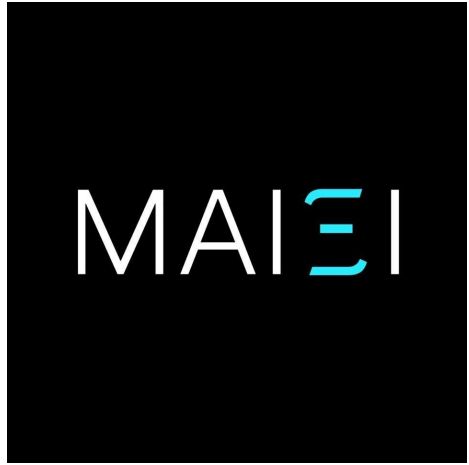

Montreal AI Ethics Institute

*An international, non-profit research institute helping humanity define its place in a world increasingly driven and characterized by algorithms*

Website: https://montrealethics.ai
Newsletter: https://aiethics.substack.com

# Report prepared by the Montreal AI Ethics Institute
# In Response to Mila's Proposal for a Contact Tracing App

*Based on insights and analysis from the Montreal AI Ethics Institute (MAIEI) Staff in response to Mila's COVI White Paper*

Contacts for the report:

Allison Cohen (allison@montrealethics.ai)
Researcher, Montreal AI Ethics Institute
Consultant, AI Global

Abhishek Gupta (abhishek@montrealethics.ai)
Founder, Montreal AI Ethics Institute
Machine Learning Engineer, Microsoft

## Introduction

Contact tracing has grown in popularity as a promising solution to the COVID-19 pandemic. The benefits of automated contact tracing are two-fold. Contact tracing promises to reduce the number of infections by being able to: 1) systematically identify all of those that have been in contact with someone who has had COVID; and, 2) ensure those that *have* been exposed to the virus do not unknowingly infect others.

"COVI" is the name of a recent contact tracing app developed by Mila and was proposed to help combat COVID-19 in Canada. The app was designed to inform each individual of their relative risk of being infected with the virus, which Mila claimed would empower citizens to make informed decisions about their movement and allow for a data-driven approach to public health policy; all the while ensuring data is safeguarded from governments, companies, and individuals.

This article will provide a critical response to Mila's COVI White Paper. Specifically, this article will discuss: the extent to which diversity has been considered in the design of the app, assumptions surrounding users' interaction with the app and the app's utility, as well as unanswered questions surrounding transparency, accountability, and security.

We see this as an opportunity to supplement the excellent risk analysis done by the COVI team to surface insights that can be applied to other contact- and proximity-tracing apps that are being developed and deployed across the world. Our hope is that, through a meaningful dialogue, we can ultimately help organizations develop better solutions that respect the fundamental rights and values of the communities these solutions are meant to serve.

## Inclusion and Community Engagement

The authors of the COVI White Paper highlighted the importance of designing the contact tracing app to suit a diversity of individuals and communities, as the app will require national participation. In pursuit of designing an accessible solution to suit the diverse needs of Canadians, Mila claims to have performed "population-level inclusivity audits". However, there are some concerns about the extent to which these audits meaningfully engaged communities.

To begin with, the population-level audits were not designed in consultation with community members. Therefore, the digital tool that Mila's staff used to obtain the community's feedback may not have received the genuine input it sought as it assumed a base level of comfort with the tool. Furthermore, although Mila mentioned working with civil society organizations (CSO), there is no information as to which organizations Mila consulted. It is important for the civil society organizations to be directly cited in order for the reader to feel confident that community engagement was meaningfully conducted. Also, transparency into the CSOs that were consulted is a great way to showcase to community members that an inclusive design process was adopted and that their needs were addressed. This generates a higher level of trust from users and, thus, more sustained, widespread, and continuous use.

Furthermore, Mila does not describe how it will present users with the option of sharing their data with Mila's Machine Learning (ML) models. Depending on the messaging Mila chooses, certain user groups may feel coerced into sharing their data. Furthermore, depending on Mila's messaging, failure to share data with the developer's ML models could lead to stigmatization among members of the community. Therefore, it is important for Mila to demonstrate how it intends to present the option to users.

In addition, "sexual orientation" is not included as a marginalized community with which Mila has consulted. However, this community may have a unique perspective on COVI and may be reluctant to share information about their location for a variety of reasons. If Mila made the explicit decision to not include the LGBTQ+ community in consultation, it is recommended that Mila make that decision known. To this end, we propose a participatory design approach for achieving some of the stated goals as specified in the COVI White Paper. MAIEI researchers Abhishek Gupta and Tania De Gasperis have shared their work on participatory design applied to contact- and proximity-tracing apps, which can be found [here](#).

Finally, the use of aggregated anonymized data, which is being sent to the government to inform health policy, risks providing unnecessary insight into citizens' movement and behavior. Also, it is unclear how transparent the government must be when making decisions informed by this data. This is especially concerning since many communities may not be engaged by the app and therefore will be underrepresented in the data and thus, the policies.

**Assumptions Surrounding Intended Use**

The COVI White Paper has implicit assumptions about how users will interact with the COVI app. However, some of those assumptions may result in poor UX design or unwarranted confidence in COVI's positive impact.

The COVI app asks users to input information about their habits, including how consistently they are washing their hands and wearing masks. However, evidence suggests that people aren't good at recording ongoing behavior. Mila should describe its decision to not use a passive data input model, which may have captured greater amounts of data than the user input model.

Furthermore, one of the most important data entry points are COVID-19 test results. However, individuals who are not technically savvy may find it difficult to submit this data point, particularly the elderly. Thus, problems of accessibility may impact the accuracy of the app. Not to mention, authentication mechanisms, designed to validate the information with health authorities, will place an additional burden on those who are new to high-technology use, limiting the widespread entry of official diagnosis into the system.

In addition, the paper mentions, "it is enough that a small fraction of the population consent to sharing their data for the ML models to already greatly enhance what is currently feasible in terms of epidemiological understanding and forecasting under different public health policies". It would be useful for Mila to provide a threshold of users needed to achieve a better understanding of epidemiological trends because it is unclear that, in the event that very few people use the app, the expected results would *justify* the risks of adopting an AI-powered contact-tracing app. Furthermore, it is unclear whether Mila considered the ethical concern about fairness if only a segment of the population needs to sacrifice their data for the rest of the community to benefit from heightened awareness about the spread. Also, Mila should provide greater detail into whether a limited sample of users can serve to provide meaningful insight into a diversity of epidemiological modeling and understanding.

A key element in establishing whether the app is fair is that its use is "*proportional* to the seriousness of the public health threat"[1]. However, in certain parts of Canada, such as the territories and Maritime provinces, the virus is *not* enough of a threat to justify using this app. It is unclear whether Mila will push for adoption of the app in communities where the risk of data breach is not proportional to the seriousness of the public health threat.

Also, since individuals are notified the day after they've been exposed to the coronavirus, it seems the individual would be reasonably capable of determining who had given them the virus, especially in sparsely populated areas. This presents data and privacy risks and potentially exposes individuals to liability if they had been alerted of their high-risk status before putting others at risk. This accountability may undermine COVI's claim that the use of the app will not expose individuals to criminal liability. Furthermore, this risk could entail government involvement to an extent that is worth mentioning in the White Paper.

In addition, Mila's decision to have COVI wait a full day before notifying an individual of their high-risk status (even when this information is readily available) risks putting other individuals at risk.

Finally, it is important for this app to recognize its potential in setting a dangerous precedent. Specifically, COVI might pave the way for contact tracing to become the default solution to future social problems. This point should be addressed so as to avoid future unnecessary or rash use of contact tracing solutions.

## Assumptions Surrounding Efficacy

In the COVI White Paper, Mila boasts the app's capacity to provide meaningful data-driven insights to policy makers and accurate risk estimates to users. Furthermore, the app claims to create a significant change in the user's behavior based on the app's recommendations. However, to what degree can these assumptions be established?

---

[1] Morley, Jessica and Cowls, Josh and Taddeo, Mariarosaria and Floridi, Luciano, Ethical Guidelines for SARS-CoV-2 Digital Tracking and Tracing Systems (April 22, 2020). Available at SSRN: https://ssrn.com/abstract=3582550 or http://dx.doi.org/10.2139/ssrn.3582550

First, Mila claims to have trained its ML algorithm using synthetically-generated data. However, it is assumed that the synthetic data that Mila used sufficiently represents the gamut of communities and individuals who will be using this app. It would be useful to explicitly describe the process through which Mila generated this synthetic training data and to what extent it can be relied upon as the basis of informing its algorithms. Given that this synthetic data, and simulations based on this synthetic data, are used to make claims about the efficacy of an ML-backed solution, transparency is critical for both the code and data used.

Mila also claims that COVI will be capable of reducing the transmission of the virus. However, Mila does not provide the data it used to inform these claims. This data should be provided in order for the graphs about Mila's relative impact to be independently verified. Without this data, it is difficult to ascertain the impact COVI will have on containing the spread of COVID-19.

Furthermore, Mila claims to use a centralized machine learning (ML) algorithm to track epidemiological trends, which will be useful for policy-makers. However, it is not clear whether the pseudonymized data being fed into the centralized ML algorithm will be tested for robustness to ensure predictions are being made on a sufficient number of data points. If this can only be done once a critical mass of users is reached, what does this imply for app's recommendations before reaching that critical mass? Will early adopters put too much trust in the algorithm dictating their behavior? If the problem of robust datasets is not addressed, it is possible that the trends observed by the ML will be inconsistent with the trends observed in reality. This also has the unfortunate consequence of potentially misleading policy makers, which can lead to disastrous consequences. Furthermore, the technical backing behind this solution risks "mathwashing" and leading people to trust the system without undertaking the necessary due diligence.

Second, Mila's COVI team does not appear to have tested the impact of its recommendations on users' behavior. The ability for the app to produce behavioral change is critical to its overall success. Therefore, demonstrating COVI's capacity to be trusted by users who are willing to act on the app's suggestions is vital and should not be assumed. In the graphs that highlight the ability of COVI to bring the $R\_0$ below 1 (i.e. limiting the pandemic from spreading), solid empirical evidence must be provided in order for people nationwide to adopt the solution.

It's also important to consider the assumption that users will self-report. It is conceivable that users may under-report their symptoms and hence skew the algorithm's output. How has Mila accounted for inaccurate data points and their ability to skew the algorithm's findings? Not to mention, accounting for the risk that poor "data-driven" policy recommendations can have when they are acted upon by health authorities.

Finally, it is unclear whether Mila considered the possibility that COVI's "high risk" prediction will skew its own models. Will COVI be tracking its false positives and provide feedback to its ML algorithms to enhance accuracy? Failure to do so could have serious implications for the algorithm's understanding of the epidemiology of the virus.

A publicly monitorable dashboard that shares essential statistical elements like the confusion matrix will help evoke trust from users, something that COVI recognizes as a key aspect of its success.

## Transparency

The COVI White Paper makes explicit mention of many concerns pertaining to a contact tracing app. However, Mila does not provide sufficient transparency into COVI's design or management decisions. Clarity into the coding, design, and management decisions is necessary for audits to be conducted before the technology becomes unleashed on the public.

The White Paper mentions a "progressive disclosure" model to balance user experience with system transparency. The progressive disclosure model begins by providing users with a clear and easy-to-understand graphic, telling the user which data is being taken by COVI. More detailed and nuanced information about user's data is presented progressively based on user literacy and interest. However, it is important for Mila to be transparent in the design trade-offs that were made in its progressive disclosure model. These decisions must be independently tested to ensure the trade offs do not undermine the goal of educating users about the data they are consenting to share. Otherwise, the model may put those who do not seek more information at risk of providing uninformed consent.

The code for the app, along with the data used to train COVI's simulators, have also not been made available yet. In order for this code to be independently audited and for Mila's assertions to be credibly validated, a third party must have access to this information, which Mila does not appear to be providing. We strongly advocate for an open-source, open-access model, especially as it relates to the tried-and-tested model from cybersecurity whereby having more eyes evaluate a solution helps to surface blindspots and build more robust and secure coding.

In addition, in conducting the privacy enhancing k-anonymity method, the developers did not clarify how they determined the level of k they sought to achieve. Will the level of k be dynamic based on the dissemination area and the composition of people's habits and behaviors? Information surrounding this method is critical to ensure Mila's privacy protections are effective.

Mila *does* mention that they have met SOC2 compliance measures, a quality assurance check on features including: security, confidentiality, privacy, availability, and processing integrity. However, the COVI White Paper does not describe how those metrics were assessed to meet the SOC2 compliance thresholds. Without this transparency, it is unclear whether meeting the SOC2 compliance measures will meaningfully increase user's trust.

Finally, without releasing COVI's code or the data used to train COVI's ML algorithm, the benefits of this app cannot be ascertained, which may lead this solution to be falsely advertised as effective. Not to mention, adverse consequences may arise if the app is ineffective to the point of telling users they are low risk when they are in fact high risk, giving them a false sense of confidence and subsequently enabling them to engage in risky behavior.

## Accountability

In order for public trust to be garnered, it is important that structures of accountability be well established in the COVI White Paper. However, COVI's accountability has not been well detailed.

Mila describes the involvement of several bodies to oversee the epidemiological insights revealed through this app. However, it is unclear whether there is a strict separation between the developers, managers, and the non-profit tasked with assessing the insights of COVI's ML system. In order for there to be accountability between these bodies, they must be separate[2]. Not to mention, it is important for the public to know who is responsible for the design, management, and oversight to ensure no conflict of interest exists.

Mila has also not provided any details into the body or individual that can be held to account if the app does not perform as intended, with consequent damages to users. Theoretically, would the Canadian government be to blame for giving the app a verbal endorsement or is it Mila for housing the creators and developers of the COVI team?

Without accountability there is no assurance that COVI's malfunctioning or damages to users will be compensated or meaningfully addressed.

## Security

The COVI White Paper mentions the risk of apps being designed to look like COVI in order to mislead users into providing their data. However, Mila does not provide information about how this problem could be remediated to compensate users.

Furthermore, although details are provided into how the system can be compromised from a cybersecurity infrastructure perspective, there is no deeper investigation into the system's exposure to adversarial ML designed to compromise the security of user's data.

Thus, COVI's security concerns have not been entirely addressed by the White Paper.

## Conclusion

Overall, Mila's White Paper had an impressive breadth and scope and has pre-emptively shone a light onto the potential risk areas that a contact tracing solution will face. However, MAIEI would like to see more details on the questions raised in this report. We look forward to engaging in a dialogue with Mila, other contact tracing app developers as well as concerned citizens and organizations to build a solution that truly benefits all Canadians.

---

[2] Alsdurf, Hannah et al.,COVI White Paper - Version 1.0  (May 18, 2020). pg 54.  Available at: https://arxiv.org/pdf/2005.08502v1.pdf